\documentstyle[aps]{revtex}
\begin{document}
\title{Localizations in coupled electronic chains}
\author{Hiroyuki Mori}
\address{Department of Quantum Matter Science,
ADSM, Hiroshima University,
Hiroshima 739-8526, Japan}
\maketitle
%
%
\begin{abstract}
We studied effects of random potentials
and roles of electron-electron interactions
in the gapless phase of coupled Hubbard chains,
using a renormalization group
technique.  For non-interacting electrons, we obtained
the localization length proportional
to the number of chains, as already shown in the other
approaches.
For interacting electrons, the localization length
is longer for stronger interactions,
that is, the interactions counteract the random potentials.
Accordingly, the localization length is not a simple
linear function of the number of chains.
This interaction effect is strongest when
there is only a single chain.
We also calculate the effects of interactions
and random potentials on charge stiffness.
\end{abstract}
\tighten
%
%
\section{Introduction}
One of the attractive topics in mesoscopic one
dimensional (1D) electron systems is the disorder effect.
While no extended state can survive in the presence
of random potentials in 1D infinite systems, 
the systems of finite size can be metallic if the size is
smaller than the localization length.
It is always important to take a serious look at
interplay between random potentials and electron-electron
interactions when we discuss the transport properties of
mesoscopic metallic wires, which are now available
as an achievement of the recent technological progress. 
An interesting example is a mesoscopic metallic ring,
which have attracted much attention
since a large persistent current was observed even with
a modest amount of impurities in the ring \cite{levy}.
A simple study, only taking an account of the impurity effect,
failed to explain such large current \cite{altshuler},
and it is necessary to
consider electron-electron interactions.

It is in general quite difficult to take correctly account
of interactions. One of the exceptions is 1D systems,
where interactions can be treated rather rigorously
with help of the bosonization techniques.
A renormalization group (RG) calculation for 1D
systems with impurities was performed
by Giamarchi and Schulz \cite{giamarchi2}.
One of their interesting results is that
partilce-particle interactions of spinless Fermions
would enhance the disorder effect and
help the system localize, while the interactions
of spinning Fermions ({\it e.g.} Hubbard type interactions)
would counteract the disorder \cite{giamarchi1}.

The RG calculation showed that the effective backward
impurity scattering $W$ of a spinless Fermion system
is renormalized as
$dW/dl=(3-2K_{\rho})W$
where $K_{\rho}$ is the Luttinger parameter of
the charge mode \cite{giamarchi2}.
The larger repulsive interactions
give rise to the smaller $K_{\rho}$
and therefore to the stronger effective impurity potentials.
This is due to the enhancement of charge density wave (CDW)
correlations in the ground state of the system.
The repulsive interactions would enhance
the CDW correlations, which make the system easily pinned by the
impurities.
For electrons with spins, however,
the renormalized impurity potential is given by
${\rm d}W/{\rm d}l=(3-K_{\rho}-K_{\sigma}-y)W$,
where $K_{\nu}\ \ (\nu=\rho, \sigma)$ is the Luttinger parameters
of charge and spin modes respectively and $y$ measures
the backward scattering strength between electrons of opposite
spins \cite{giamarchi2,giamarchi1}.
For Hubbard model with small $U$, this RG equation becomes
\begin{equation}
\frac{dW}{dl}=\left(1-\frac{U}{\pi v_{F}}\right)W,
\label{W}
\end{equation}
where $v_F$ is the Fermi velocity.
Here the electron-electron interactions
would screen the impurity potentials.
Since a spin density wave (SDW) correlation is dominant
in the ground state of the repulsive
Hubbard model, the interaction $U$ makes
the particle density uniform,
and therefore make the coupling of the density to the impurities
weak.

These opposite roles of interactions
in the spinless and spining Fermions
was later checked numerically in connection
with the problem of persistent current \cite{kato,mori,morihamada}.
Although the suppression of the effective disorder strength
due to electron-electron interactions in the models
with the spin degrees of freedom might
explain the observed large persistent current in the
disordered metal rings, we should carefully consider the fact
that the metal rings have the finite cross section
and therefore have the finite number of channels.

Ladder systems have recently attracted a wide range of attentions.
Disorder effect on electronic ladder systems
is however not widely
investigated so far, and there are few studies
which calculated RG equations for two-leg electronic ladder
with impurity potentials \cite{orignac,fujimoto,mori2}.
If we just focus on the role of interactions
in the presence of impurities, it is shown that
the role changes drastically depending on the phase;
the impurity effect is enhanced by electron-electron
interactions in C1S2 phase (CnSm means that n charge modes and
m spin modes are gapless) \cite{orignac}
while it is suppressed 
in C2S2 phase (with all modes being gapless) \cite{mori2}.
Remember that the ground state of a single chain is gapless 
and the impurity effect is suppressed by the interactions.
So, as far as we concern the gapless phase, the role
of interactions acting against random potentials
does not alter even when we increase the number of channels
from one to two. But the point we should note is
that the suppression of the effective disorder strength
due to interactions is smaller in two channel ladder
than single channel chain \cite{mori2}.
If this tendency continues in the systems with more
channels, we could not expect the large interaction-driven
suppression of electron localizations in a thin but finite
cross-section wire. It is therefore of importance
to see in the increased number of channels
whether electron-electron interactions counteract
random potentials and if so how large the effect is.

In this paper we first show in Sec. 2 how the disorder effect
on a non-interacting electronic ladder
changes as the number of chains is increased.
In Sec. 3 we turn on electron-electron interactions
and investigate the interplay between the interactions
and the impurity potentials in one-, two-, and three-chain ladders.
Section 4 is devoted for summary of the paper.
%
%
\section{Noninteracting disordered chains}
We start from a tight-binding Hamiltonian of coupled chains with
open boundary condition,
\begin{equation}
H_0=-t\sum c^{\dagger}_{m\sigma i}c_{m\sigma i+1}
-t_{\perp}\sum c^{\dagger}_{m\sigma i}c_{m+1\sigma i}
+h.c.,
\label{Horiginal}
\end{equation}
where $m$ and $i$ are the chain and site indices,
respectively. This hopping
terms are diagonalized by a unitary transformation
$c_m=\sum_{\alpha}V_{m\alpha}a_{\alpha}$ and they form bands.
The matrix $V$ for one, two, three-chain systems,
for example, is given by
\begin{equation}
V=\left\{
\begin{array}{ll}
1&\hspace*{1cm}(N_{ch}=1),\\
&\\
\left(
\begin{array}{cc}
\frac{1}{\sqrt{2}}&\frac{1}{\sqrt{2}}\\
\frac{1}{\sqrt{2}}&-\frac{1}{\sqrt{2}}
\end{array}
\right)
&\hspace*{1cm}(N_{ch}=2),\\
&\\
\left(
\begin{array}{ccc}
\frac{1}{2}&\frac{1}{\sqrt{2}}&\frac{1}{2}\\
\frac{1}{\sqrt{2}}&0&-\frac{1}{\sqrt{2}}\\
\frac{1}{2}&-\frac{1}{\sqrt{2}}&\frac{1}{2}
\end{array}
\right)
&\hspace*{1cm}(N_{ch}=3),
\end{array}
\right.
\end{equation}
where $N_{ch}$ is the number of the chains.
Following the standard bosonization procedure, we get
the Tomonaga-Luttinger type Hamiltonian
\begin{equation}H_0=\sum_{\nu \alpha}\int\frac{dx}
{2\pi}[u_{\alpha\nu}K_{\alpha\nu}(\pi\Pi_{\alpha\nu})^2+
\frac{u_{\alpha\nu}}{K_{\alpha\nu}}(\partial_x\phi_{\alpha\nu})^2],
\end{equation}
where $\nu=\rho,\sigma$ represents charge and spin
mode respectively, and $\alpha$ is the band index.
With no interaction, the parameters $K$ are equal to
1 and $u$ are 
the Fermi velocities of each band.

The impurity scattering potential term is originally
written in the form
\begin{eqnarray}
H_{imp}&=&\sum n^{imp}_{m i}c^{\dagger}_{m\sigma i}
c_{m\sigma i}\\
&=&\sum n_{mi}^{imp}V^*_{m\alpha}V_{m\beta}
a^{\dagger}_{\alpha\sigma i}a_{\beta\sigma i},
\end{eqnarray}
where $V^*$ is the complex conjugate of $V$.
After the bosonization we get the forward and backward
scattering terms.
Since the forward scattering does not affect $H_0$\cite{fujimoto},
we only consider the effect of the backward impurity scattering.
The backward scattering terms are given by
\begin{equation}
H_b=
\frac{1}{\pi a}\int dx\sum_{\alpha \beta ,s}
\xi_{\alpha\beta}e^{-i\{\phi_{\alpha\rho}+
\phi_{\beta\rho}+s(\phi_{\alpha\sigma}+\phi_{\beta\sigma}
)\}/\sqrt{2}}
\cos[\{(\theta_{\alpha\rho}-\theta_{\beta\rho})
+s(\theta_{\alpha\sigma}-\theta_{\beta\sigma})\}/\sqrt{2}]
+h.c,
\end{equation}
where $a$ is the lattice constant.
$\xi$ is assumed to be a linear
combination of Gaussian random variables and then the
following general form of the action will appear
in the replica trick method by integrating
out the random variables,
\begin{eqnarray}
S_b^{imp}=
&&\frac{2}{(\pi a)^2}\sum_{\alpha\beta\gamma\delta}
\sum_{ijss'}Z^{\alpha\beta}_{\gamma\delta}\int
\cos\{(\theta^i_{\alpha\rho}-\theta^i_{\beta\rho}
+s(\theta^i_{\alpha\sigma}-\theta^i_{\beta\sigma}))
/\sqrt{2}\}
\cos\{(\theta^j_{\gamma\rho}-\theta^j_{\delta\rho}
+s'(\theta^j_{\gamma\sigma}-\theta^j_{\delta\sigma}))
/\sqrt{2}\}
\label{Sb}
\\
&&\times
\cos [\{(\phi^i_{\alpha\rho}+\phi^i_{\beta\rho}+
s(\phi^i_{\alpha\sigma}+\phi^i_{\beta\sigma}))
-(\phi^j_{\gamma\rho}+\phi^j_{\delta\rho}+
s'(\phi^j_{\gamma\sigma}+\phi^j_{\delta\sigma}))\}
/\sqrt{2}],\nonumber
\end{eqnarray}
where $i,j$ are the replica indices.
In order to discuss the disorder effect of various coupled
chains on the same basis, we assume $n^{imp}_{mi}$
is a Gaussian random variable and satisfies
$<n^{imp}_{mi}n^{imp}_{lj}>=W\delta_{ml}\delta_{ij}$.
Hence we have
\begin{equation}
Z^{\alpha\beta}_{\gamma\delta}=
W\sum_nV^*_{n\alpha}V_{n\beta}V_{n\gamma}V^*_{n\delta}.
\label{Z}
\end{equation}

The RG equation for $Z^{\alpha\beta}_{\gamma\delta}$ is
\begin{equation}
\frac{dZ^{\alpha\beta}_{\gamma\delta}}{dl}
=Z^{\alpha\beta}_{\gamma\delta},
\label{RGZ1}
\end{equation}
where we used the fact that
$K$'s are all equal to 1 for free particles.
The solution of the RG equation is
$Z^{\alpha\beta}_{\gamma\delta}(l)=%
e^lW\sum_nV^*_{n\alpha}V_{n\beta}V_{n\gamma}V^*_{n\delta}$.
Note that $We^l=\sum_{\beta} Z^{\alpha\beta}_{\alpha\beta}(l)$.
The RG equation stops when $Z^{\alpha\beta}_{\gamma\delta}(l)$
reaches $\sim \overline{v_F}^2/a$,
and it is when $l\sim\log(L_{loc} /a)$ where $L_{loc}$
is the localization length. Putting all together, we get
\begin{equation}
L_{loc}\sim N_{ch}\frac{\overline{v_F}^2}{W}.
\end{equation}
On the other hand, the electron scattering rate is given by
$\tau ^{-1}\sim W\rho (\epsilon_F)$, where
$\rho (\epsilon_F)$ ($\sim \overline{v_F^{-1}}$) is the density
of states at the Fermi level, and hence the mean free
path $l_e=\overline{v_F}\tau \sim\overline{v_F}^2/W$. Then we get
the known relation \cite{pichard,tamura},
$L_{loc}\sim N_{ch}l_e$.

In order to see the disorder effect on the transport
properties, we next calculate charge stiffness $D$,
which measures the strength of Drude peak,
$\sigma =D\delta (\omega )+\sigma_{reg}$.
Note that $D$ is also a measure of the persistent
currents for small flux.

The external flux couples with the current
$\sum_{\alpha s}j_{\alpha s}$, where $\alpha$ is
the band index. 
Since $\sum_{\alpha s}j_{\alpha s}\propto%
\sum_{\alpha}\Pi_{\alpha\rho}$,
charge stiffness is proportional
to $\sum_{\alpha}K_{\alpha\rho}u_{\alpha\rho}$.
Ignoring the irrelevant numerical factor,
we define charge stiffness per channel by
$D=(1/N_{ch})\sum_{\alpha}K_{\alpha\rho}u_{\alpha\rho}$.
For non-interacting and non-impurity ladder,
$D$ is the averaged Fermi velocity $\overline{v_F}$.

In the presence of impurities
the RG equation for the charge stiffness is given by
\begin{eqnarray}
\frac{dD}{dl}&=&-\sum_{\alpha}\frac{2a
u_{\alpha\rho}^2} {u_{\alpha\sigma}^3\pi}
(\frac{1}{N_{ch}}\sum_{\beta}Z^{\alpha\beta}_{\alpha\beta})
\\\nonumber
&=&-\frac{2a}{\pi}\overline{v_{F}^{-1}}We^l.
\label{RGD}
\end{eqnarray}
$D$ has no explicit $N_{ch}$ dependence and the effect
of increasing channel number could appear only through
$\overline{v_{F}^{-1}}$, which is usually weak.
For a finite system of size $L$, $D$ is given by
\begin{equation}
D\sim D_0-\frac{2}{\pi}\overline{v_F^{-1}}WL,
\label{D}
\end{equation}
where $D_0$ is a constant. For a given size $L$,
the charge stiffness $D$ is smaller for stronger
disorder, and for a given disorder $W$, $D$ is smaller
for a larger system. The former is trivial, and the latter
is because $L_{loc}/L$ becomes smaller as $L$ increases
with $W$ fixed and hence the system will be in
more localized regime. This will be clear if we rewrite
Eq.(\ref{D}) as
\begin{equation}
D\sim D_0-\frac{2}{\pi}N_{ch}\overline{v_F}\frac{L}{L_{loc}}.
\end{equation}

We cannot simply extend
these discussion to $N_{ch}\rightarrow\infty$, that is,
to two dimensions (2D), because we constructed the RG theory 
based on the Tomonaga-Luttinger type bosonized Hamiltonian
with well-defined subbands,
which is not an appropriate basis for 2D system.
%
%
\section{Interacting disordered chains}
In the previous section we showed that the localization
length is proportional to the number of chains,
and the effect of random potentials becomes
weaker as the chain number increases. What happens
if we turn on interactions between the particles.
The original Hamiltonian is then Eq.(\ref{Horiginal})
plus the Hubbard-type interaction term,
\begin{equation}
U\sum n_{m\uparrow i}n_{m\downarrow i}.
\end{equation}
A part of the interaction terms can be combined
with the kinetic term to give
\begin{equation}H_0=\sum_{r\nu}\int\frac{dx}
{2\pi}[u_{r\nu}K_{r\nu}(\pi\Pi_{r\nu})^2+
\frac{u_{r\nu}}{K_{r\nu}}(\partial_x\phi_{r\nu})^2],
\end{equation}
where $\nu=\rho,\sigma$ represents charge and spin
mode respectively, and $r$ is the new band index
assigned by the unitary transformation
$\phi_{\alpha\nu}\rightarrow\sum_r%
T_{\alpha r}\phi_{r\nu}$ and
$\Pi_{\alpha\nu}\rightarrow\sum_r%
T_{\alpha r}\Pi_{r\nu}$, where 
\begin{equation}
T=\left\{
\begin{array}{ll}
1&\hspace*{1cm}(N_{ch}=1),\\
&\\
\left(
\begin{array}{cc}
\frac{1}{\sqrt{2}}&\frac{1}{\sqrt{2}}\\
\frac{1}{\sqrt{2}}&-\frac{1}{\sqrt{2}}
\end{array}
\right)
&\hspace*{1cm}(N_{ch}=2),\\
&\\
\left(
\begin{array}{ccc}
\frac{1}{\sqrt{2}}&\frac{1}{\sqrt{3}}&\frac{1}{\sqrt{6}}\\
0&\frac{1}{\sqrt{3}}&-\sqrt{\frac{2}{3}}\\
-\frac{1}{\sqrt{2}}&\frac{1}{\sqrt{3}}&\frac{1}{\sqrt{6}}
\end{array}
\right)
&\hspace*{1cm}(N_{ch}=3).
\end{array}
\right.
\end{equation}
This unitary transformation is necessary to diagonalize $H_0$.
Since a part of interaction terms are
included in $H_0$, the parameters $K$ and $u$ have different
values from the ones of the noninteracting case.
Hereafter we use Greek letters
$\alpha ,\beta ,\cdots$ for the old band index and
Roman letters $r, \cdots$ for the new band index.

The other remaining interaction terms are represented by $H_1$.
Since there are still a lot of terms in $H_1$, we do not
write them down all here and we only note that the interaction
matrix element $g_{\alpha\beta\gamma\delta}$ in the commonly
used notation,
representing the scattering from $(\alpha , \beta )$
bands to $(\gamma , \delta )$ bands, is given by
\begin{equation}
g_{\alpha\beta\gamma\delta}=
U\sum_lV_{l\alpha}V_{l\beta}V_{l\gamma}^*V_{l\delta}^*
\end{equation}
in the case of Hubbard model,
although some of them, such as $g_{\alpha\beta\beta\alpha}$ and
$g_{\alpha\beta\beta\beta}$, should be set to zero
due to lack of the momentum conservation.
As stressed in Ref. \cite{schulz2},
one should not forget the sign factors of $g$ which
come from Majorana Fermion operators.
In the present case, the backward scattering interaction
$g^{(1)}_{\alpha\alpha\beta\beta}$, for example,
has to be multiplied by $-1$.

The presence of the interaction terms made the phase diagram
quite rich\cite{schulz2,fabrizio,balents,nagaosa,noack,%
endres,asai,yamaji,kuroki}, and the interplay between interactions
and random potentials is strongly phase dependent as
stated in Sec. 1.
In order to construct a systematic view with changing the number
of chains, we should pick up a common phase,
{\em gapless} phase, where no gap opens both in the charge
and spin modes.
There is also another reason to study gapless phase.
When the forward impurity scatterings are considered,
Fujimoto and Kawakami showed \cite{fujimoto} that
the electron-electron interactions are effectively suppressed
and as a result the charge and spin gaps collapse.
In this situation the remaining task is to study
the effect of the backward impurity scatterings on the
gap-collapsed (gapless) phase.

In the gapless phase the action presenting the backward
impurity scatterings is again given by Eq.(\ref{Sb}).
The RG equation for $Z^{\alpha\beta}_{\gamma\delta}$ is
\begin{eqnarray}
\frac{dZ^{\alpha\beta}_{\gamma\delta}}{dl}
&=&[3-\frac{1}{8}\sum_{r}\{\{
(T_{\alpha r}-T_{\beta r})^2
+(T_{\gamma r}-T_{\delta r})^2\}(K^{-1}_{r\rho}
+K^{-1}_{r\sigma})\nonumber\\
&&+\{
(T_{\alpha r}+T_{\beta r})^2
+(T_{\gamma r}+T_{\delta r})^2\}(K_{ r\rho}
+K_{r\sigma})\}]Z^{\alpha\beta}_{\gamma\delta}
\label{RGZ2}\\
&&-\frac{1}{u_{\sigma}\pi}\sum_{\alpha'\beta'}
g^{(1)}_{\alpha'\gamma\delta\beta'}
Z^{\alpha\beta}_{\alpha'\beta'},\nonumber
\end{eqnarray}
where $g^{(1)}$ represents the backward scattering
interactions. Note that, for noninteracting electrons, 
we have $K=1$ and $g=0$ and then Eq. (\ref{RGZ2})
reduces to Eq. (\ref{RGZ1}). Also, when $N_{ch}=1$,
Eq. (\ref{RGZ2}) reduces to $dW/dl=(1-U/u_{\sigma}\pi )W$
which is just what we mentioned in Eq. (\ref{W}).

Since the renormalization of
$Z^{\alpha\beta}_{\gamma\delta}$
has a strong $K$ dependence, the localization length
and other physical quantities would changes
in accordance with $K$ in a complicated way.
To make the story simple,
we consider the weak interaction limit and use
$K=1$ as in the noninteracting system. Also we assume
the random impurity potentials are so weak that
we can forget about the renormalization of $K$.
The RG equation (\ref{RGZ2}) can then be written in
the following matrix form,
\begin{equation}
\frac{dZ}{dl}=(1-G)Z.
\label{RGZ4}
\end{equation}
The $(i,j)$ element of the $N_{ch}\times N_{ch}$
matrix $Z$ is
$Z^{\alpha\beta}_{\gamma\delta}$ where
$i=(\alpha, \beta )$ and $j=(\gamma ,\delta )$.
The $(i,j)$ element of the matrix $G$
is given by 
$\frac{1}{u_{\sigma}\pi}g^{(1)}_{\gamma\alpha\beta\delta }$
where $g^{(1)}$ is now assumed to include
the sign created by the Majorana Fermion operators
mentioned above. The signs of the  elements of $G$ 
are not always plus and it is not in general trivial whether
the presence of $G$ in Eq. (\ref{RGZ4}) would
weaken the growth of $Z$, namely, whether the interactions
would suppress the impurity effects.
So let us see it in more detail.

Since we assume the interactions are weak and
ignore the renormalization of $G$, the solution
of Eq. (\ref{RGZ4}) is
$Z(l)=e^{(1-G)l}Z(0)\sim e^l(1-Gl)Z(0)$.
Therefore $Tr(Z(l))\sim e^l\{N_{ch}W-lTr(GZ(0))\}%
\sim N_{ch}We^{(1-CU/\overline{v_F}\pi )l}$,
where $C=u_{\sigma}\pi Tr(GZ(0))/(N_{ch}WU)$.
Since, when $l\sim \log (L_{loc}/a)$,
$Tr(Z(l))\sim N_{ch}^2\overline{v_F}^2/a$,
we have $N_{ch}\overline{v_F}^2/Wa\sim %
(L_{loc}/a)^{1-CU/\overline{v_F}\pi }$.
Then the localization length of
the interacting system has the form,
\begin{equation}
\frac{L_{loc}}{a}\sim \left(\frac{L_{loc}^{(0)}}{a}\right)
^{1+CU/\overline{v_F}\pi},
\label{loc}
\end{equation}
where $L_{loc}^{(0)}=N_{ch}\overline{v_F}^2/W$ is the
localization length of the noninteracting system.
Because of the presence of $U$, the localization
length is no longer simply proportional to $N_{ch}$.
If we forget about the sign of $g_{\alpha\beta\gamma\delta}$
created by the Majorana Fermion algebra, discussed above,
$C=1$ for any $N_{ch}$. Taking the sign into account, however,
$C$ becomes
\begin{equation}
C=\left\{
\begin{array}{ll}
1&\hspace*{1cm}(N_{ch}=1),\\
0.25&\hspace*{1cm}(N_{ch}=2),\\
0.4\cdots &\hspace*{1cm}(N_{ch}=3).
\end{array}
\right.
\end{equation}
Therefore the interaction $U$ always makes the 
localization length longer, namely, the interactions have
a delocalization effect in the gapless phase
of the coupled chain systems.
$C$ is not a monotonic function of $N_{ch}$ as far as
$1\leq N_{ch}\leq 3$.
and hence it is not always true that the interaction
effect becomes weaker as $N_{ch}$ increases.
Anyway, since it is easy to prove $C\leq 1$,
the delocalization effect is strongest when $N_{ch}=1$.

Next we study the charge stiffness $D$. As noted
in the previous section, the external flux couples to
$\sum_{\alpha s}j_{\alpha s}\propto\sum_{\alpha}%
\Pi_{\alpha\rho}=\sum_{\alpha r}T_{\alpha r}\Pi_{r\rho}$.
Recall that the Greek letter $\alpha$ represents
the old band index assigned before the operation of $T$,
and the Roman letter $r$ is the new band index.
Since $\sum_{\alpha}T_{\alpha r}=1$ (when $N_{ch}=1$),
$\sqrt{2}\delta_{r1}$ (when $N_{ch}=2$), $\sqrt{3}\delta_{r2}$
(when $N_{ch}=3$), the charge stiffness is given by
$D=u_{r\rho}K_{r\rho}$ where $r=1$ (when $N_{ch}=1,2$),
$r=2$ (when $N_{ch}=3$) with the irrelevant numerical
factor being omitted. The RG equation is then given by
\begin{eqnarray}
\frac{dD}{dl}&=&-\frac{2au_{r\rho}^2}
{u_{r\sigma}^3\pi}\sum_{\alpha\beta}Z^{\alpha\beta}
_{\alpha\beta}T_{\alpha r}^2
\label{RGD3}\\
&\sim &-\frac{2a}{\pi}\overline{v_F}^{-1}
We^{(1-CU/\overline{v_F}\pi )l},\nonumber
\end{eqnarray}
where $r=1$ (when $N_{ch}=1,2$), $r=2$
(when $N_{ch}=3$). In the second line of Eq. (\ref{RGD3})
the interactions are assumed to be small.
For a finite system of size $L$,
\begin{equation}
D\sim D_0-\frac{2a}{\pi}\overline{v_F}^{-1}W\left(\frac{L}{a}\right)
^{1-CU/\overline{v_F}\pi}.
\end{equation}
Again, the interaction $U$ suppresses the random potential effect
and gives larger value of the charge stiffness that in the
noninteracting case.
%
%
\section{Summary}
We investigated the role of electron-electron interactions
in the coupled Hubbard chains with random potentials.
For noninteracting systems, a RG calculation shows that the 
effective strength of the impurity potentials grows
towards the strong coupling limit and the localization
length is proportional to the number of chains $N_{ch}$.
In the presence of interactions between the particles,
the role of the interactions changes from the phase
to phase as previously shown in Ref. \cite{orignac,mori2},
and therefore we only focused on the gapless phase of
$N_{ch}$ chain systems. Based on the RG calculation,
we showed that the interactions always reduce the effective
strength of the impurity potentials. The degree of
the reduction is sensitive to $N_{ch}$,
and the counter-effect of the interactions against the
random potentials is strongest when $N_{ch}=1$.
%
%

\end{document}